\documentclass[times, 10pt,twocolumn]{article} 
\usepackage{latex8}
\usepackage{times}

\usepackage[dvips]{epsfig}
\usepackage{color}
\usepackage{hyperref}
\usepackage[htt]{hyphenat}

\begin{document}

\title{Power Law Distributions in Class Relationships}

\author{
	Richard Wheeldon and Steve Counsell\\
  School of Computer Science and Information Systems\\
  Birkbeck College, University of London\\
  London WC1E 7HX, U.K.\\
  \{richard, steve\}@dcs.bbk.ac.uk\\
}

\date{\today}
\maketitle

\begin{abstract}
Power law distributions have been found in many natural and social phenomena,
and more recently in the source code and run-time characteristics of
Object-Oriented (OO) systems. A power law implies that small values are
extremely common, whereas large values are extremely rare.
In this paper, we identify twelve new power laws
relating to the static graph structures of Java programs. The graph structures
analyzed represented different forms of OO coupling, namely, inheritance,
aggregation, interface, parameter type and return type. Identification
of these new laws provide the basis for predicting likely features of classes
in future developments. The research in this paper ties together work in object-based
coupling and World Wide Web structures.
\end{abstract}

\section{Introduction}

Power law distributions have been found in many natural and social phenomena.
A power law implies that small values are extremely common, whereas large
values are extremely rare.
For example, incomes, earthquake strengths, city sizes and word frequency all
follow power law distributions - there are many small tremors, but only a few
large earthquakes.
The power law distribution is strongly connected with Zipfs-law and the Pareto
distribution, often known as the 80:20 rule \cite{ADAM00b}.

We would expect a power law to apply to the size of classes in object-oriented
systems. Size in this sense is defined in terms of the number of methods,
constructors and other class features.
This hypothesis is partially supported by previous research into
{\em key} classes \cite{COUN00}.

The existence of a power law distribution in a network implies a \emph{scale-free}
behaviour. This means that the network lacks a ``characteristic length scale'' so that,
like fractals, when suitably magnified, small bits of it resemble the whole.
Whichever range of values is examined, the proportion of small to large
values remains the same \cite{ADAM00b}.

Recently, there has been a great deal of interest in the power laws visible in
the structure of the World Wide Web \cite{ADAM02}. These include the number
pages on web sites, the number of links to a given website, the PageRank
\footnote{ The ranking metric used by Google \cite{PAGE98} }
values of nodes and the frequency with which users visit pages
\cite{PANG02,ADAM02}.

The {\em indegrees} and {\em outdegrees} of individual web pages are also subject
to power law distibutions \cite{BROD00b}.
In a Web context, indegree refers to number of pages linking to a given page
and outdegree refers to the number of pages referenced from a given page.
During the development of our Autodoc system for assisted navigation of program
documentation \cite{WHEE02}, we found that the pages in the Javadocs were subject
to the same laws. We thus hypothesized that the relationships were due to power
law distributions in the underlying code structure.

Our motivation for the research described in this paper is to discover patterns
and relationships which can explain the structure of source code at a
low level of abstraction. Identifying such patterns allows us to predict, by
extrapolation, the consequences of developing larger and more complex software.
For example, we could predict how many classes might contain greater than a
hundred methods in a set of classes ten times larger than the Java Developers
Kit (JDK). Alternatively, we could predict the maximum number of constructors
of any class in that system. This may have implications for software
maintenance and comprehensibility in terms of time spent and effort
expended \cite{NAJJ03}.

The identification of power law distributions in source code enables us
to categorize the coupling graphs as belonging to a class of scale-free
topologies. Similar topologies have been found in the Web \cite{BROD00b},
the Internet \cite{FALO99} and in some food webs \cite{GARL03}. Techniques
to store, manipulate and describe such topologies may be re-applied to handle
coupling graphs.

A further motivation is to enable models of code development that will allow
developers to create synthetic code bases containing large numbers of computer-generated
classes. For example, given an appropriate means of generating synthetic data,
a developer could generate a data set of a much larger number of classes.
This would enable them to test the consequences of developing a large system
before development begins.

Finally, our work has implications for the graph traversal algorithms used in
reachability analysis and garbage collection. Just as
internet networks are robust against random removal of nodes \cite{ALBE00}, it is
likely that random removal of classes will have little effect on the proportion of
code which can be reached and thus executed.

The remainder of this paper is organized as follows. Section~\ref{sec:related} describes
related work in discovering power law distributions and scale-free characteristics
of software. In Section~\ref{sec:experimentalmethods} we describe our analysis techniques
and present the results in Section~\ref{sec:results}. Section~\ref{sec:conclusion} gives
our conclusions and ideas for future research.

\section{Related Work}
\label{sec:related}

There has been substantial work on power law distributions in natural phenomena
and over recent years, in the evolution of the web. Only recently has attention
turned to the power law distributions found in program code and, in particular,
those relating to Java software.

O'Donoghue et al. have performed a run-time analysis of Java bytecode sequences
obtained using a customized version of the Kaffe JVM \cite{ODON02}. Their
experiments showed that the frequencies with which consecutive pairs of
instructions are interpreted by the virtual machine follows a power law.

Potanin et al. have conducted experiments using a query-based analysis tool
called Fox, which is an enhanced version of Bill Foote's Heap Analysis Tool (HAT).
Their research has confirmed power laws in the indegree and outdegrees of the
run-time object graphs of several programs \cite{POTA02b,POTA03}.

Valverde et al. have shown that the indegree and outdegrees of nodes in a
network of class diagrams also follow power laws, leading to a scale-free network
topology similar to that of the Web \cite{VALV02}. Since these diagrams have a
one-to-one mapping with the source code structure, the implication is that these
laws are a feature of object-oriented program code.

A major feature of the work in this paper is the analysis of different coupling
types. A commonly-held view in software engineering is that there is a link between
complexity in software and the understandability of that software. The more
coupling in a system, the more complex the system. 
Too much coupling may be indicative of a poorly thought out design or of
inadequate standards of maintenance. There is evidence to suggest that excessive
coupling can lead to more fault-prone software \cite{BRIA97,HARR98}.
The large number of different ways of writing OO code means that accurately 
capturing, categorising and analyzing the different forms of coupling is a difficult 
task to undertake.
A comprehensive framework for measuring coupling in OO systems is described
by Briand et al \cite{BRIA99}.

Our research shows that even when the network of classes is decomposed by coupling
type, power laws still prevail. This has lead to the identification of twelve
distinct power law distributions.

\section{Analysis Techniques}
\label{sec:experimentalmethods}

\begin{figure*}[tbp]
	\begin{center}
	\psfig{figure=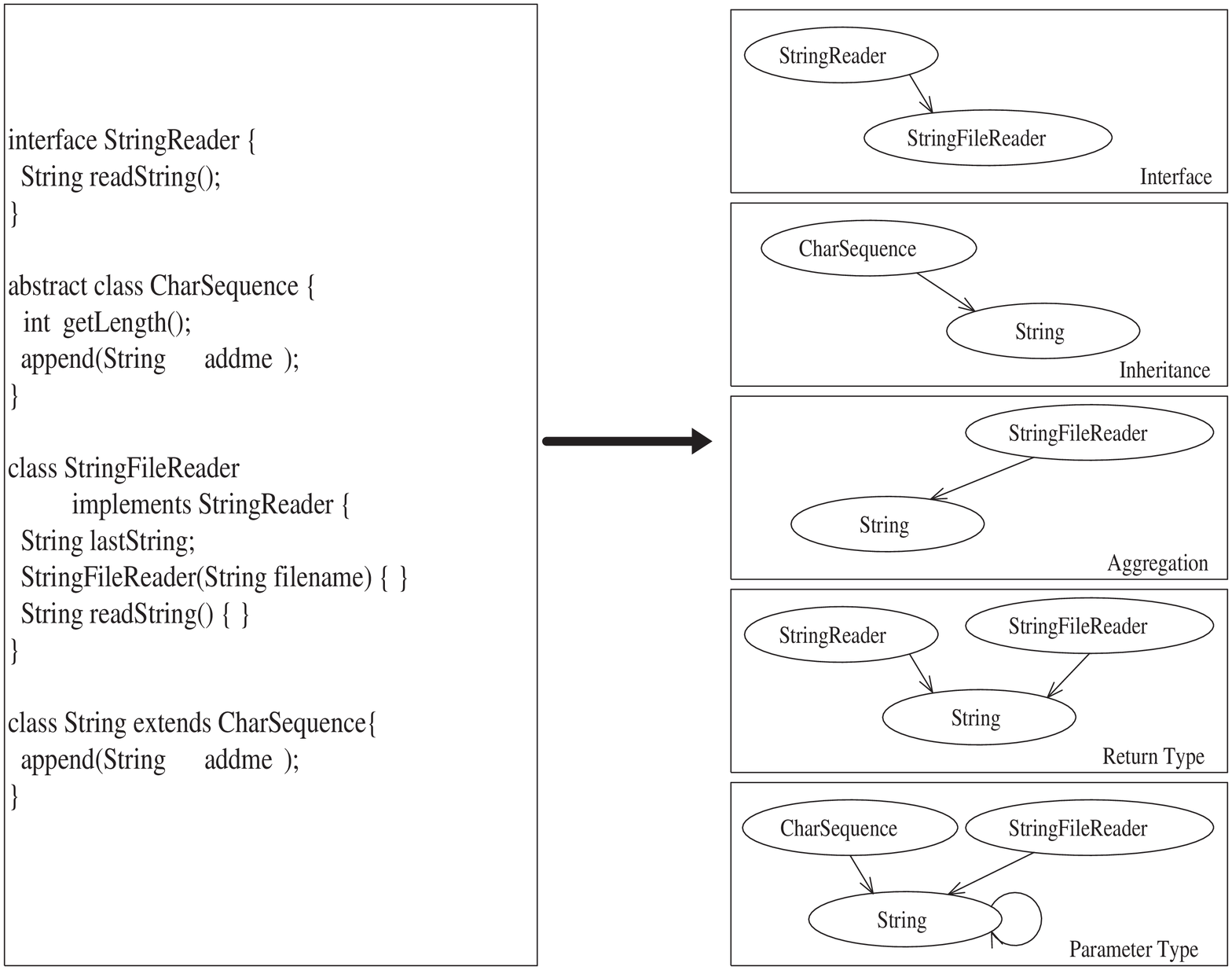, width=150mm}
	\caption{ \label{fig:CodeGraphs}
		Illustration of coupling types and their graph representations.
	}
	\end{center}
\end{figure*}

As part of this research, a system called AutoCode has been developed for indexing
Java source code. AutoCode works by using a custom {\em doclet} which extends the
Javadoc program and allows easy access to the code structure. We used the AutoCode
system to generate graphs for each of five coupling types - Inheritance, Interface,
Aggregation, Parameter Type and Return Type. An illustration of how these graphs can
be derived from source code can be seen in Figure~\ref{fig:CodeGraphs}. To identify
the power laws, we then performed statistical operations on these five graphs.
With the exception of inheritance, we need to consider the relationships from two
perspectives. For example, from the perspective of the interface and of the
implementing class. We do not need to consider the number of superclasses because
Java classes only ever have one superclass.
The number of methods, constructors and fields in each class were also studied.

Data was collected from three large Java systems:
\begin{enumerate}
\item The core Java class libraries shipped with the {\em Java Developers Kit}
	(JDK) provide implementations of common functions required for many programs.
	Version 1.4.1 of the JDK contains 1~400~000 lines of code spread over 6~000 classes.
\item {\em Apache Ant} is a Java-based build tool. It behaves in a similar way to
	\texttt{make} but uses XML-based configuration files, which define various
	tasks to be executed. The source code	for version 1.5.3 of Apache Ant
 	contains 145~000 lines of code spread over 500 classes.
\item {\em Tomcat} is the servlet container used in the official reference
	implementation for Java Servlets and JavaServer Pages. The source code for
	Jakarta Tomcat version 4.0 contains 150~000 lines spread over 370 classes.
\end{enumerate}

To identify the power laws we perform linear regression on log-log data plots. 
The number of occurrences, $y$, of
a value of magnitude $x$ is given by the equation $y = C x^{-a}$ which
implies that $\log(y) = \log(C) - a \log(x)$. Hence, the power law can easily be
identified by a straight line with gradient $-a$ on a log-log plot.
Because of significant clustering of data points near the $x$-axis, regression on
these plots leads to skewed results. To prevent this, the values must be
grouped into buckets of exponentially increasing size \cite{ADAM02}. The logarithm
of the frequency is plotted against the logarithm of the mid-point of each bucket.
From the subsequent regression a more accurate exponent value, $a$, can be obtained
than if all the original data points are considered. It is this value which allows
us to predict the likely features of future systems. A low value of the
exponent signifies a tendency towards a less skewed distribution.

\section{Results}
\label{sec:results}

\subsection{Methods, Fields and Constructors}

\begin{figure*}
   \begin{center}
        \begin{tabular}{cc}
					\psfig{figure=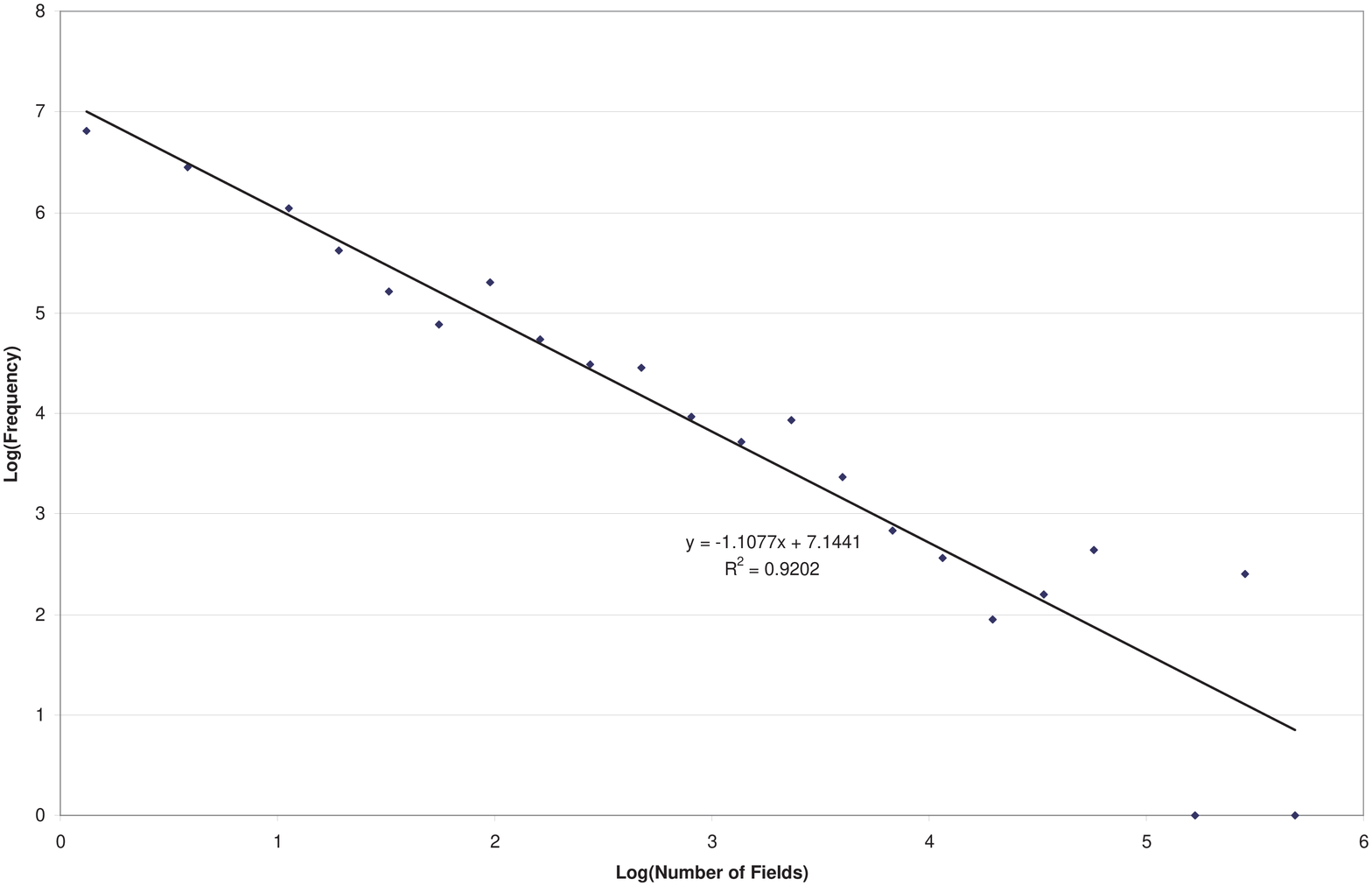, width=75mm} & 
					\psfig{figure=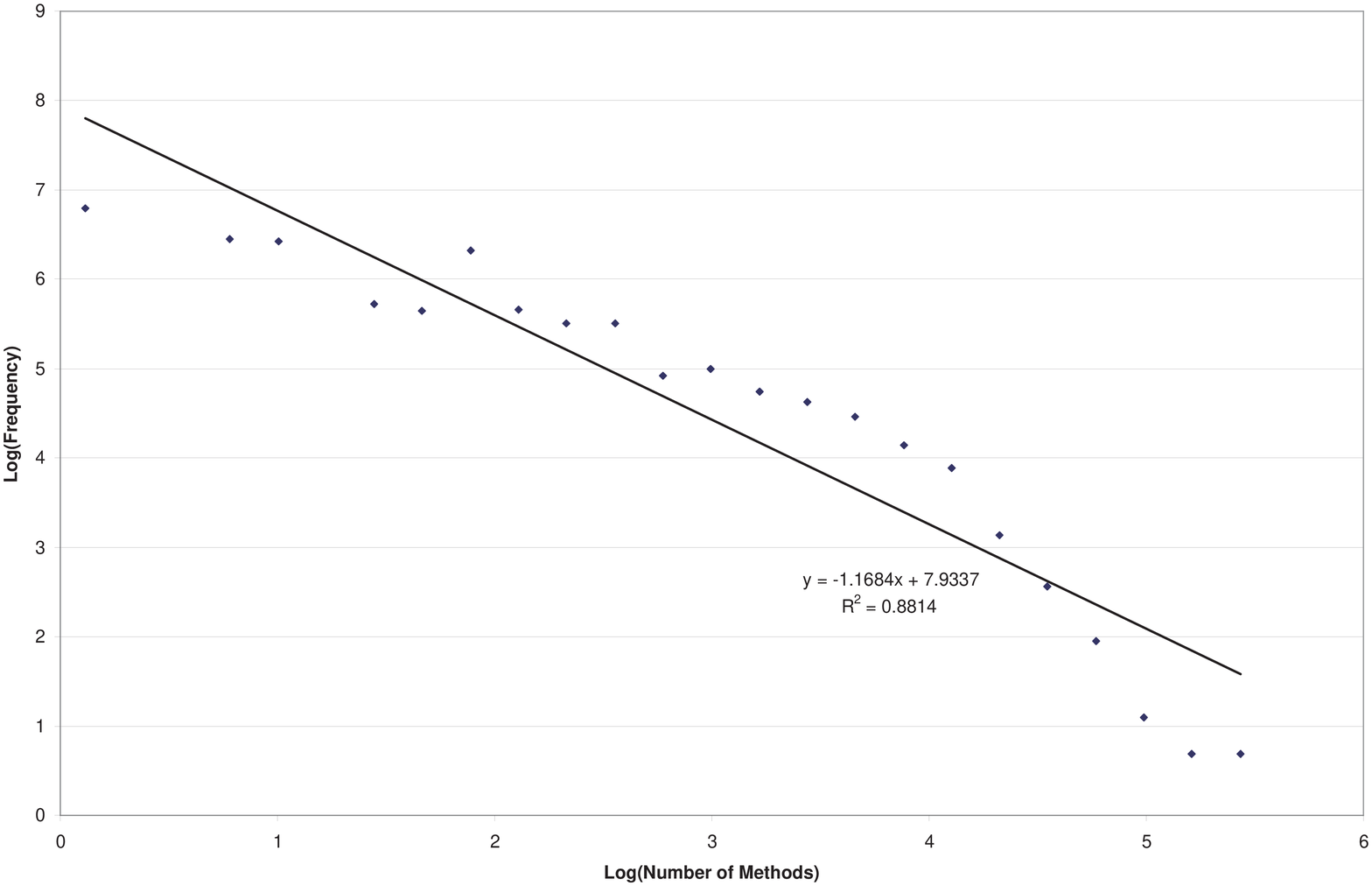, width=75mm} \\
          (a) Fields &
          (b) Methods \\
				\end{tabular}
				\begin{tabular}{c}
					\psfig{figure=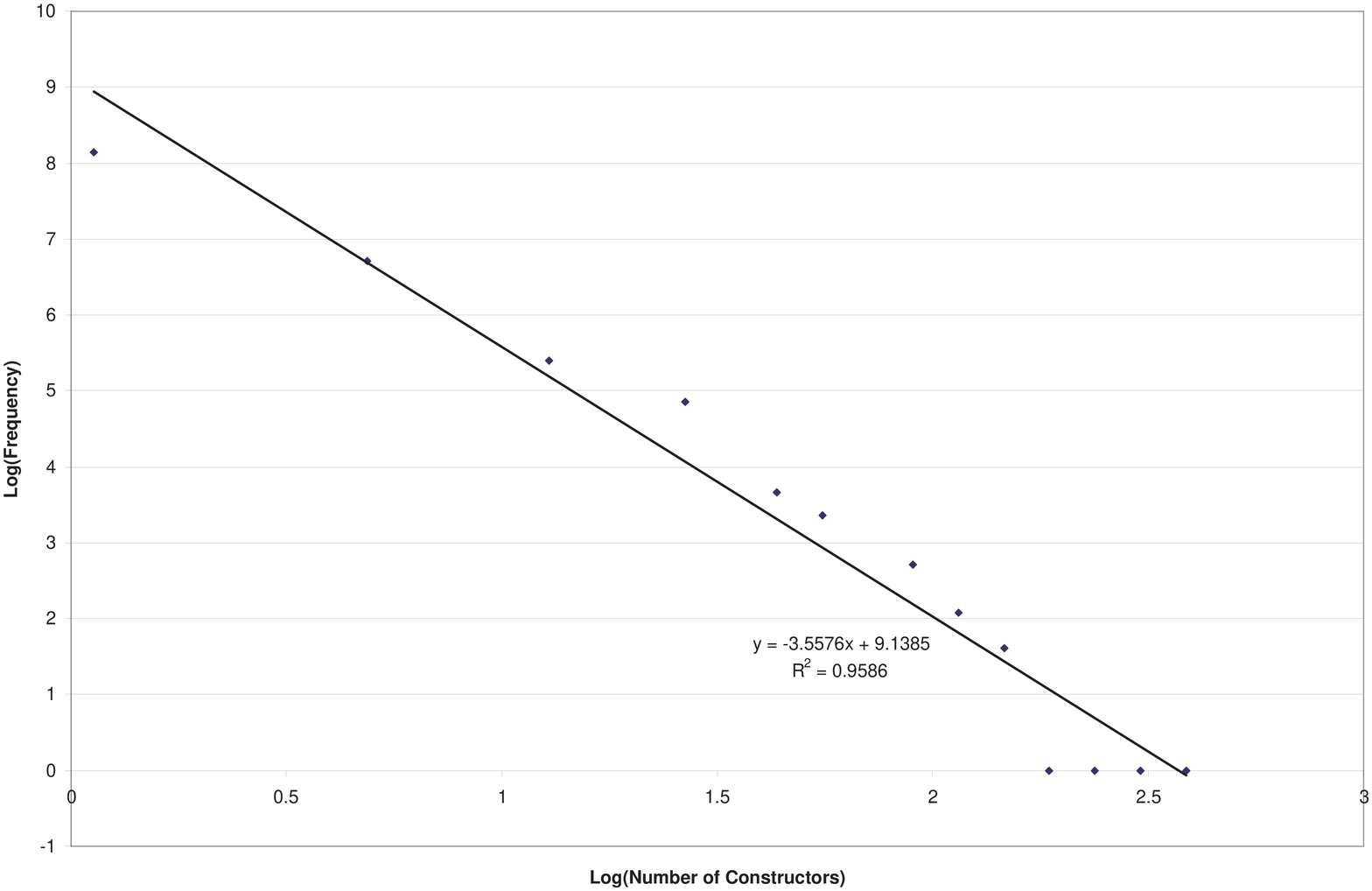, width=75mm}\\
					(c) Constructors \\
         \end{tabular}
  	\caption{\label{fig:mfc-counts}
			Log-log plots showing power law distributions in the number of (a) fields,
			(b) methods and (c) constructors of classes in the JDK class libraries.
			}
   \end{center}            
\end{figure*}

The majority of this study concerns coupling relationships between classes. However,
three power laws were identified without type information. These relate to the fundamental
building blocks of classes - the number of fields in each class, the number of methods
in each class and the number of class constructors.
Figure~\ref{fig:mfc-counts} shows log-log plots highlighting each of these relationships.

\begin{figure*}
   \begin{center}
        \begin{tabular}{cc}
						\psfig{figure=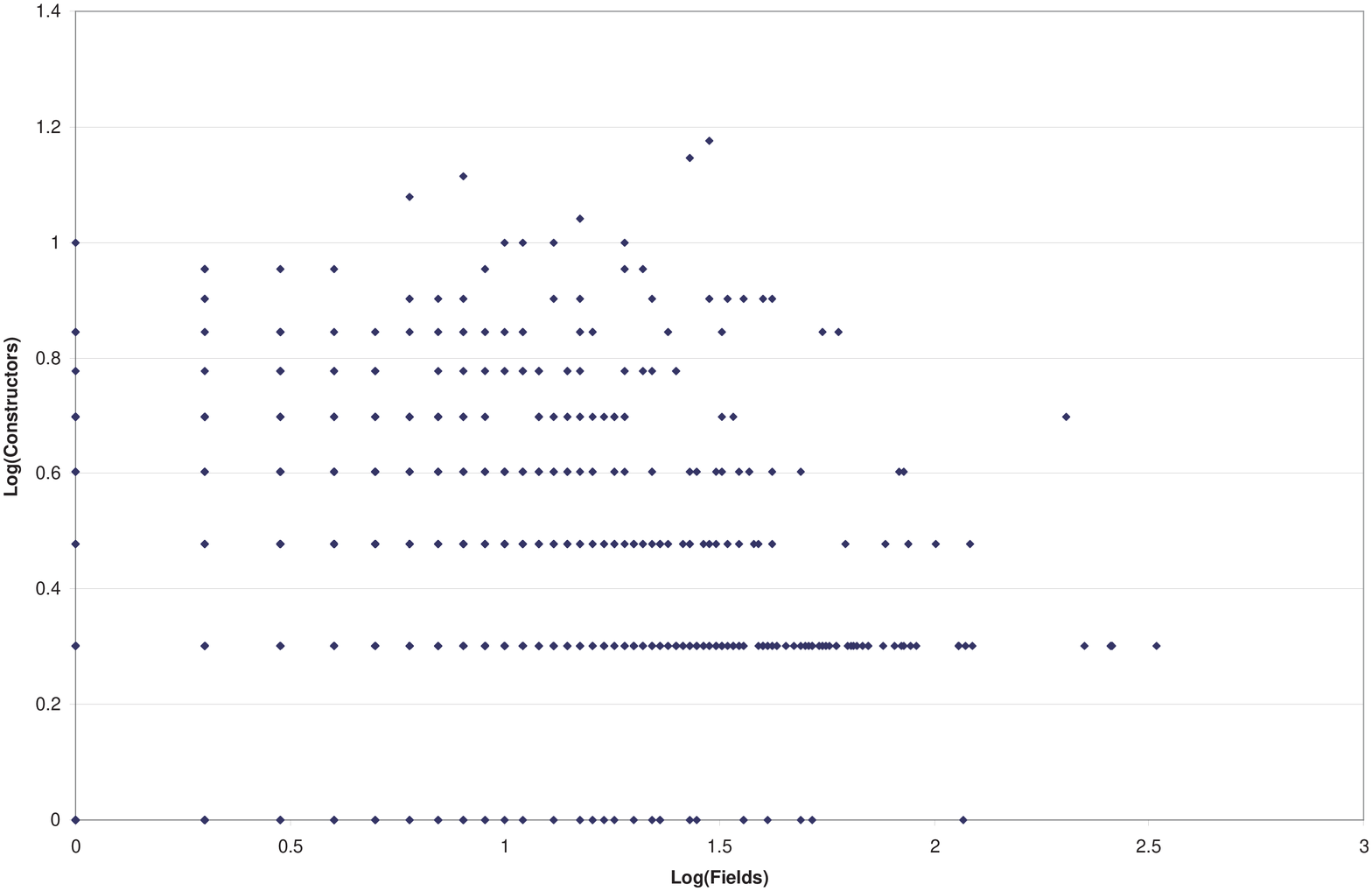,width=75mm} &
						\psfig{figure=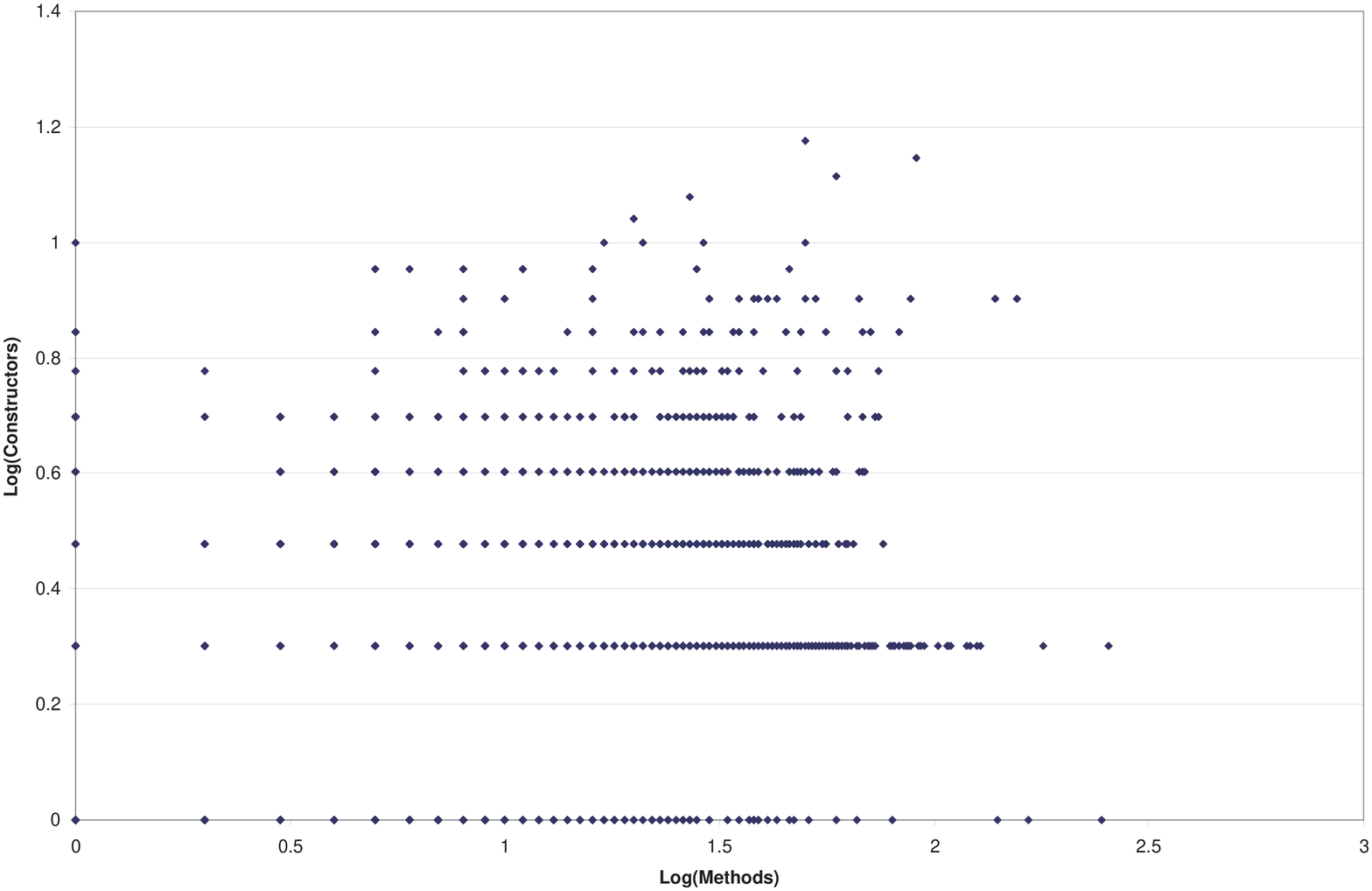,width=75mm} \\
            (a) Fields vs. Constructors &
            (b) Methods vs. Constructors \\
				\end{tabular}
				\begin{tabular}{c}
						\psfig{figure=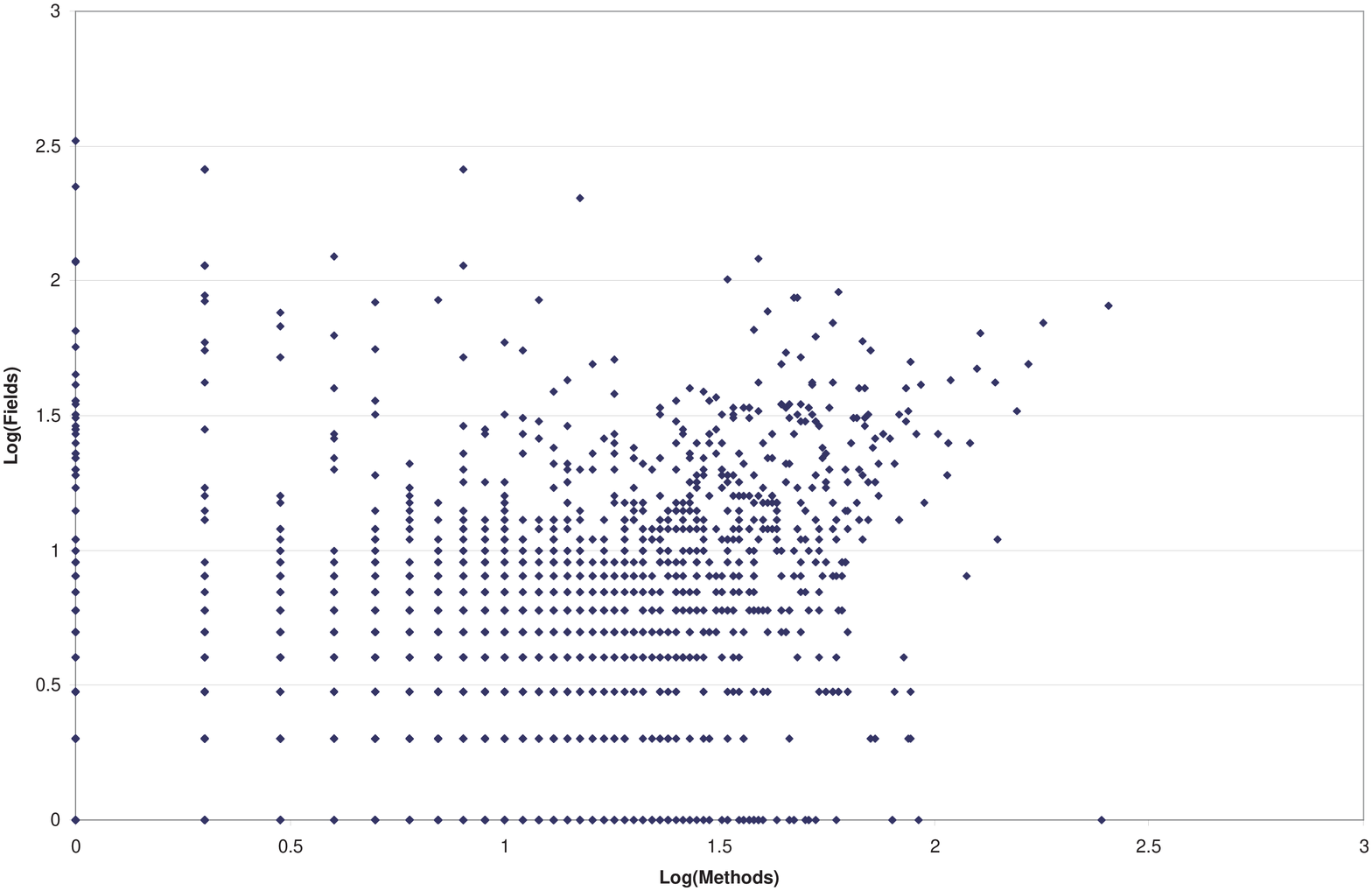,width=75mm} \\
						(c) Methods vs. Fields \\
         \end{tabular}
  	\caption{\label{fig:mfc-relationships}
			Log-log plots showing the relationships between (a) the number of fields
			and the number of constructors, (b) the number of methods and the number
			of constructors and (c) the number of methods and the number of fields
			for classes in the JDK.
			}
   \end{center}            
\end{figure*}

For the distribution of the number of methods, the exponents are 1.168, 1.104 and 0.734
for JDK, Ant and Tomcat, respectively. This implies that in the JDK there is a higher
proportion of classes with very few methods when compared with the other two systems.
This might imply fewer key classes in this system.
For the distribution of the number of fields, the
exponents are 1.108, 0.988 and 0.998 for JDK, Ant and Tomcat, respectively. The difference
in the magnitude of the exponents would indicate no strong relationship between the
number of methods and the number of fields. 
It could be imagined that a large number of fields implies a larger number of methods to
operate on those fields. Based upon our obvservation,
we hypothesize that it is infeasible to predict the number of methods from the number of
fields and vice-versa. This hypothesis is supported by correlations between the number of
methods, fields and constructors  (Figure~\ref{fig:mfc-relationships}).
The correlation matrix in Figure~\ref{fig:mfc-correlation} shows that
no strong correlation exists between any of these measures.

For the distribution of the number of constructors, the exponents are 3.560, 3.589 and 3.112
for JDK, Ant and Tomcat, respectively. This implies that classes with a large number
of constructors are rarely found in systems of this scale. For example, the JDK system
contains only three classes with more than ten constructors. Previous work into
refactoring of constructors found similar evidence for five medium-sized Java systems
\cite{NAJJ03}. Only one class was found to have ten constructors. This class was part of
the Swing library.

\begin{figure}[htbp]
\begin{center}
\begin{tabular}{|l|ccc|}\hline
	            & Methods & Fields & Constructors \\ \hline
Methods       & 1       &        & \\
Fields        & 0.216   & 1	     & \\
Constructors  & 0.215   & 0.0827 & 1 \\ \hline
\end{tabular}
\caption{
	\label{fig:mfc-correlation}
	Correlation matrix for class members in the JDK
}
\end{center}
\end{figure}

\subsection{Coupling Power-Laws}

\begin{figure}[htbp]
	\psfig{figure=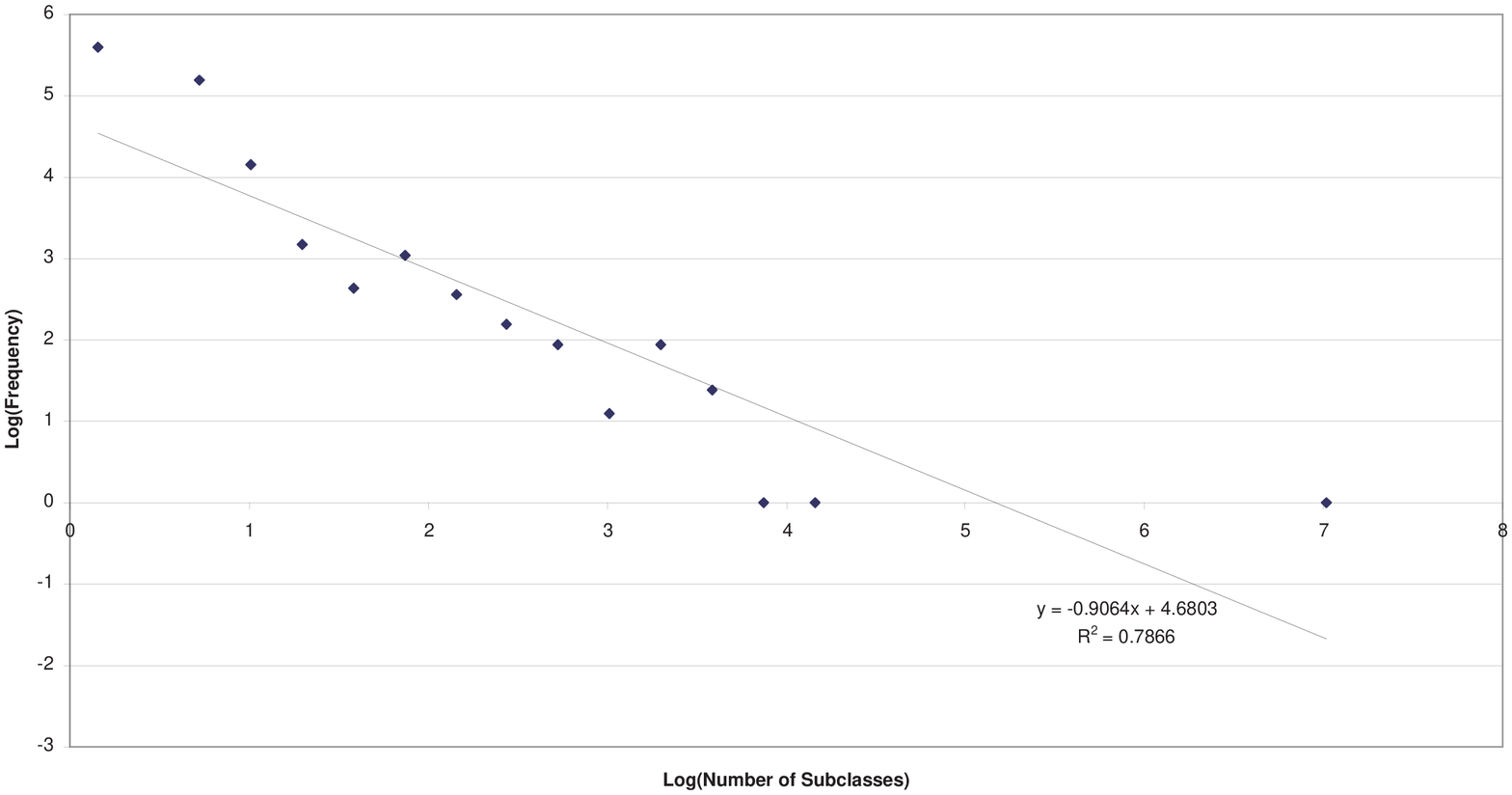, width=75mm, height=65mm}
	\caption{ \label{fig:jdk-inheritance-outdegrees}
		Log-Log plot showing a power law distribution in the number of subclasses
		of each class in the JDK class library.
	}
\end{figure}

The frequency with which classes are used as superclasses to other classes
can be calculated by examining the distribution of outlinks in the
superclass-subclass graph. Figure~\ref{fig:jdk-inheritance-outdegrees} shows a
bucketed log-log plot of the number of descendants of the classes in the JDK.
The results show that the distribution follows a power law with exponent 
0.906. The exponents for Apache Ant and Jakarta Tomcat are 0.810 and 1.310,
respectively. The high value for Tomcat implies that more classes in
that system have relatively few descendants, whilst a small number of classes 
are extended by many descendants. In other words, the functionality of the system is
distributed more evenly than in the other two systems. In contrast, for the
Ant system, much of the functionality is contained in subclasses of key classes
such as \texttt{Task} and \texttt{BaseParamFilterReader}.
Hence the functionality is more concentrated in fewer classes in this system.

By using the same techniques we can show that the distribution of
the number of classes implementing an interface follows a power law,
with exponents 1.130, 1.118 and 1.636 for JDK, Ant and Tomcat, respectively.
This makes sense if we consider the use of interfaces as a surrogate for
multiple inheritance. We would expect a similar distribution for interface
implementations as for subclasses.

\begin{figure}[htbp]
	\psfig{figure=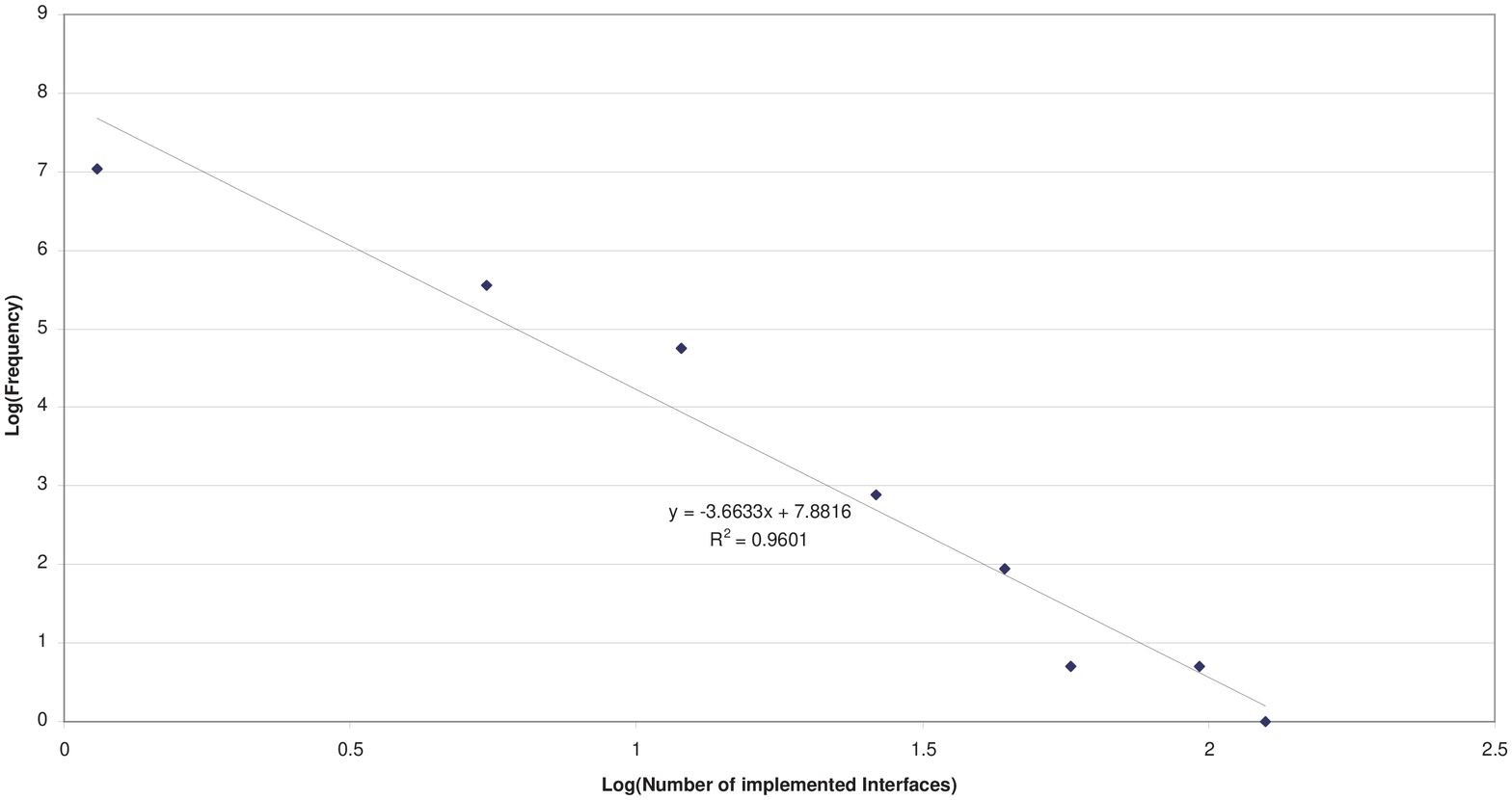, width=75mm, height=65mm}
	\caption{ \label{fig:jdk-interface-indegrees}
		Log-Log plot showing a power law distribution in the number of interfaces
		implemented by classes in the JDK class library.
	}
\end{figure}

The distribution in the number of interfaces implemented by a class also follows
a power law, with a much higher exponent of 3.663, as can be seen from
Figure~\ref{fig:jdk-interface-indegrees}. This exponent was calculated for the JDK.
Insufficient data was available to calculate the exponents for the other two systems.
This result can be explained by virtue of very few classes implementing a large
number of interfaces. Those that do implement a large number of interfaces tend to
delegate the responsibility for the methods of these interfaces to members of
the same interface.

Two further power law distributions can be seen in the relationship between classes
as member variables. The first, a power law distribution in the number of other
classes referenced as member variables within a given class.
For example, in Figure~\ref{fig:CodeGraphs}, {\tt StringFileReader} references
one class, {\tt String}, via the field {\tt lastString}. The exponents of the
distributions are 1.051, 1.386 and 1.493 for JDK, Ant and Tomcat, respectively.
The low value for JDK reflects a comparatively uniform distribution of coupling
via aggregation in this system. One explanation for the low JDK value may be
that the roles of various packages in the system do not overlap and hence
there are multiple focal points for aggregation, as opposed to a centralized
structure.

The second distribution is in the number of classes which reference a given
class as a member variable. For example, in Figure~\ref{fig:CodeGraphs},
{\tt String} is referenced by one class, {\tt StringFileReader}. The exponents
of these distributions are 1.399, 2.295 and 1.991 for JDK, Ant and Tomcat,
respectively. Interestingly the JDK again has the lowest exponent value
supporting the previous hypothesis about multiple focal points for aggregation.

\begin{figure*}
   \begin{center}
        \begin{tabular}{cc}
					\psfig{figure=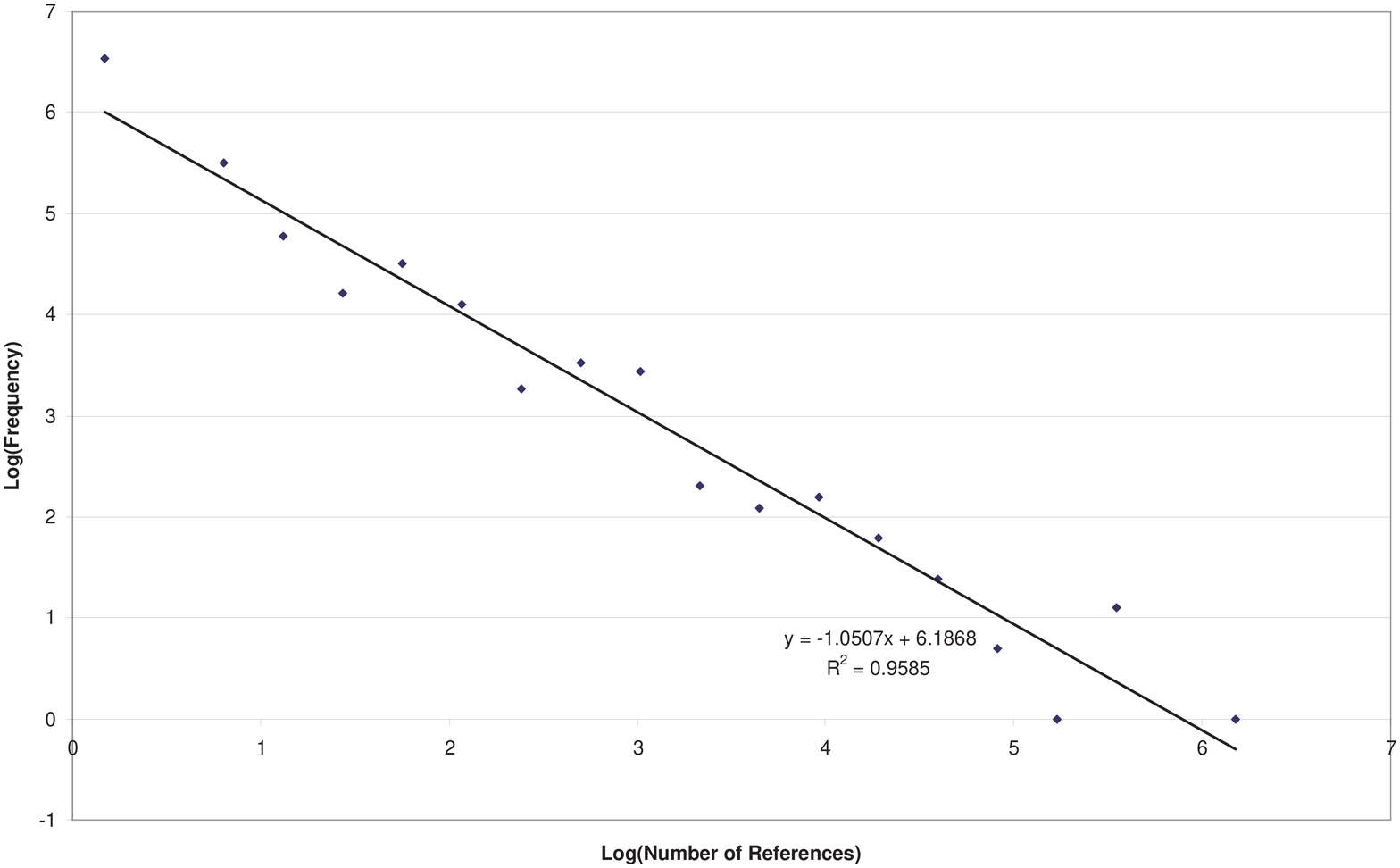, width=75mm, height=65mm} &
					\psfig{figure=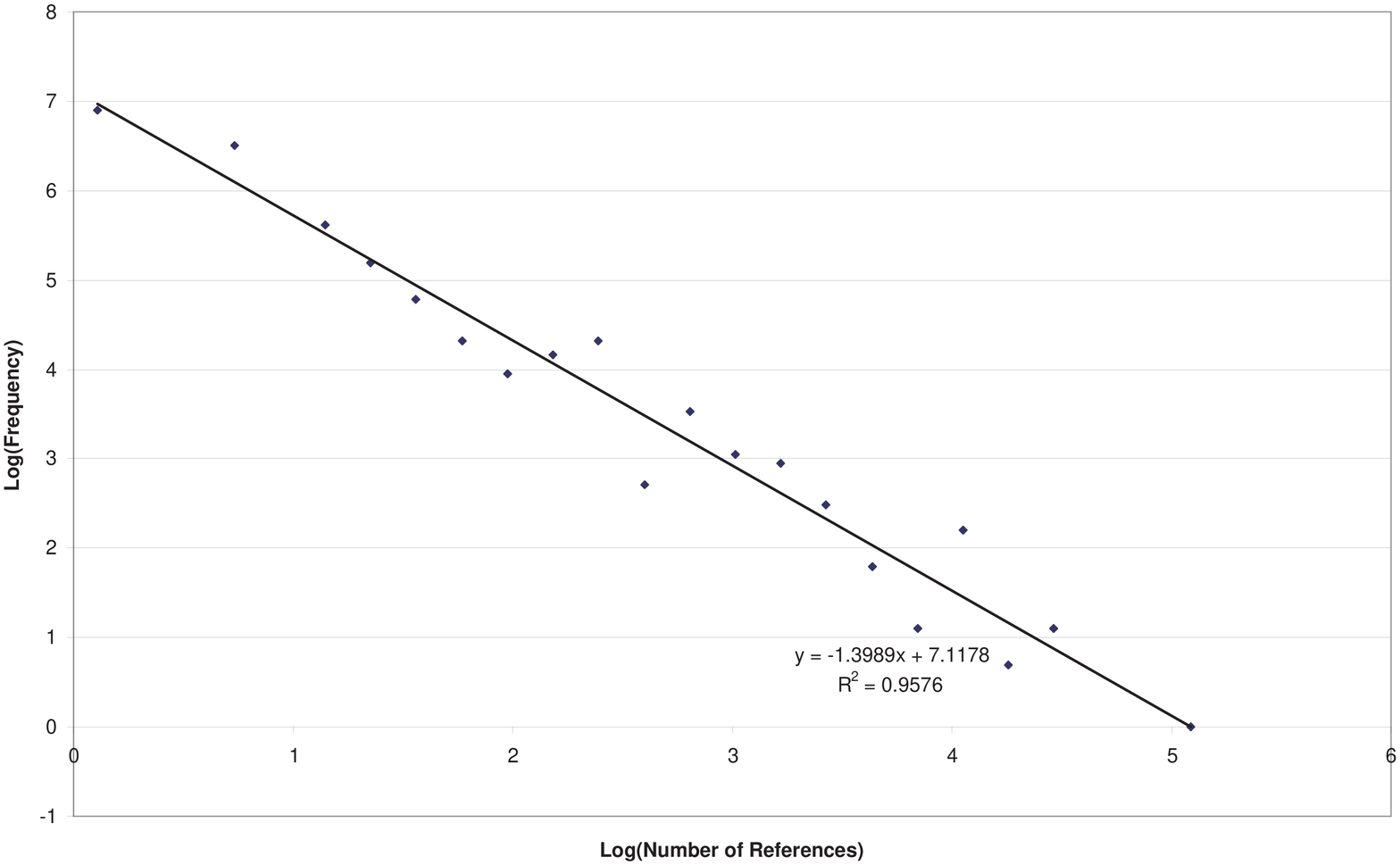, width=75mm, height=65mm} \\
          (a) Field members &
          (b) Containing classes \\
			\end{tabular}
  	\caption{ \label{fig:aggregation}
			Log-Log plots showing power law distributions in (a) the number of classes
			referenced as field variables and (b) in the number of classes which contain
			references to classes as field variables.
		}
   \end{center}            
\end{figure*}

Both of these power-laws can be seen from the plots in Figure~\ref{fig:aggregation}.
It is noticable that the values for the first distribution are lower than the
corresponding values for the second. This can be explained by the tendency in
object-oriented code for many classes to be grouped together as members of
another class. In contrast, it is comparatively rare for a class to be referenced
as a member in many classes.

Four more class features were analyzed for power law distributions, namely the
indegrees and outdegrees induced by parameter types and return types for each of
the three systems. All showed scale-free topology.
The Ant system has comparatively high values for all the exponents in these
relationships. Inspection of the classes in this system and subsequent analysis
revealed no strong correlation between usage of return types and parameters.
This could be considered a suprising result, since we might expect parameters and
return types to be linked.
No obvious explanation could be found for the differences in exponents between
the systems.

The exponent values for all three systems can be found in
figures~\ref{fig:jdk-summary},~\ref{fig:tomcat-summary}~and~\ref{fig:ant-summary}.
The $r^2$ values denote Pearson product-moment correlation.
The high $r^2$ values for JDK reflect the larger number of classes in this system.
As a result, we would expect more consistency in the data. The $r^2$ values are
relatively low but still support the theory.

\section{Conclusions and Future Work}
\label{sec:conclusion}

In this paper we have illustrated that power-law distributions exist in
object-oriented class relationships. In particular, those related to
coupling. Twelve new power-laws have been identified. The exponents of
these power laws are given for the JDK (Figure~\ref{fig:jdk-summary}),
Tomcat (Figure~\ref{fig:tomcat-summary}) and Ant (Figure~\ref{fig:ant-summary}).
One conclusion from this work is the belief that these regularities are
common across all non-trivial object-oriented programs.

Another conclusion is that the different types of coupling examined are
independent. This finding contradicts the hypothesis that high usage in one form
of coupling can be used to predict high usage in another form.

The implications of these findings are that we can use the data to predict
the dimensions of future systems. This will allow us to estimate the
complexity of developing and maintaining those systems.

It is interesting to note that the exponents for Ant and Tomcat rarely fall
within the 95\% confidence intervals of the JDK. We believe that these
exponents are due to deeper properties of the collections.
The conclusion is that whilst there
are common properties between these systems, each individual system has
its own unique characteristics.

Bieman and Murdock have already shown that there is a large body of freely
accessible source code available on the Web \cite{BIEM01}. In terms of future
work, it would be interesting to verify these results using a large crawl of
such data. Assuming that these results hold, a number of techniques can be
brought to bear to explain the phenomena.

In order to explain the power law in World Wide Web graphs, new models for its
growth and evolution have emerged. The key to these models is a process known
as \emph{preferential attachment} \cite{ALBE00b} in which pages which have a
high indegree are more likely to be referred to by new links. This can be
explained by considering a page with higher indegree as being more popular,
more important and better connected. It is thus more likely to be
visited by a user who may then also choose to link to that page. Research is
ongoing to find methods to improve the model - for example, by combining
preferential and non-preferential attachment \cite{LEVE01c}.
Other future work will investigate the accuracy with which these models
can predict the structure of program code.

\begin{figure*}[htbp]
  \begin{center}
		\begin{tabular}{|l|r|r|r|r|} \hline
Relationship & Exponent & Lower 95\% & Upper 95\% & $r^2$ \\ \hline
Number of Methods                  & 1.168 & 0.968 & 1.369 & 0.881 \\ 
Number of Fields                   & 1.108 & 0.956 & 1.260 & 0.920 \\ 
Number of Constructors             & 3.560 & 3.067 & 4.048 & 0.959 \\ 
Subclasses                         & 0.906 & 0.623 & 1.189 & 0.787 \\
Implemented Interfaces             & 3.663 & 2.918 & 4.409 & 0.960 \\
Interface Implementations          & 1.130 & 0.933 & 1.329 & 0.907 \\
References to class as a member    & 1.051 & 0.935 & 1.167 & 0.959 \\ 
Members of class type              & 1.399 & 1.253 & 1.455 & 0.958 \\ 
References to class as a parameter & 0.802 & 0.712 & 0.892 & 0.949 \\ 
Parameter-type class references    & 1.130 & 1.011 & 1.250 & 0.954 \\ 
References to class as return type & 0.914 & 0.789 & 1.039 & 0.938 \\ 
Methods returning classes          & 1.491 & 1.293 & 1.689 & 0.937 \\ 
\hline
		\end{tabular}
	\end{center}
	\caption{ \label{fig:jdk-summary}
		95\% confidence intervals for power law exponents in JDK.
	}
\end{figure*}

\begin{figure*}[htbp]
  \begin{center}
		\begin{tabular}{|l|r|r|r|r|} \hline
	Relationship & Exponent & Lower 95\% & Upper 95\% & $r^2$ \\ \hline
	Number of Methods                  & 0.734 & 0.522 & 0.965 & 0.734 \\ 
	Number of Fields                   & 0.998 & 0.775 & 1.222 & 0.858 \\ 
	Number of Constructors             & 3.112 & 2.654 & 3.570 & 0.994 \\ 
	Subclasses                         & 1.310 & 0.714 & 1.906 & 0.828 \\
	Interface Implementations          & 1.636 & 0.865 & 2.407 & 0.856 \\
	References to class as a member    & 1.493 & 1.049 & 1.937 & 0.849 \\ 
	Members of class type              & 1.991 & 1.433 & 2.550 & 0.910 \\ 
	References to class as a parameter & 0.684 & 0.368 & 1.000 & 0.569 \\ 
	Parameter-type class references    & 1.126 & 0.788 & 1.464 & 0.771 \\ 
	References to class as return type & 1.119 & 0.879 & 1.358 & 0.896 \\ 
	Methods returning classes          & 1.400 & 0.933 & 1.866 & 0.799 \\ 
 \hline
		\end{tabular}
	\end{center}
	\caption{ \label{fig:tomcat-summary}
		95\% confidence intervals for power law exponents in Tomcat.
	}
\end{figure*}

\begin{figure*}[htbp]
  \begin{center}
		\begin{tabular}{|l|r|r|r|r|} \hline
Relationship & Exponent & Lower 95\% & Upper 95\% & $r^2$ \\ \hline
Number of Methods                  & 1.104 & 0.862 & 1.345 & 0.828 \\ 
Number of Fields                   & 0.988 & 0.833 & 1.143 & 0.920 \\ 
Number of Constructors             & 3.589 & 2.722 & 4.456 & 0.958 \\ 
Subclasses                         & 0.810 & 0.452 & 1.169 & 0.667 \\
Interface Implementations          & 1.118 & 0.585 & 1.652 & 0.814 \\
References to class as a member    & 1.386 & 0.869 & 1.903 & 0.803 \\ 
Members of class type              & 2.295 & 1.951 & 2.639 & 0.967 \\ 
References to class as a parameter & 1.034 & 0.678 & 1.389 & 0.752 \\ 
Parameter-type class references    & 1.456 & 1.021 & 1.892 & 0.800 \\ 
References to class as return type & 1.001 & 0.538 & 1.463 & 0.699 \\ 
Methods returning classes          & 1.850 & 1.360 & 2.340 & 0.919 \\ 
\hline
		\end{tabular}
	\end{center}
	\caption{ \label{fig:ant-summary}
		95\% confidence intervals for power law exponents in Ant.
	}
\end{figure*}

\bibliographystyle{plain}
\bibliography{../bib/papers}

\end{document}